\documentclass[conference]{IEEEtran}
\IEEEoverridecommandlockouts
\usepackage{amsmath,amssymb,amsfonts}
\usepackage{algorithmic}
\usepackage{graphicx}
\usepackage{textcomp}
\usepackage{xcolor}
\usepackage{hyperref}
\usepackage{cite}
\usepackage{bbm}
\usepackage{subcaption}
\usepackage{mathtools}
\usepackage{bm}
\usepackage[mathscr]{euscript}
\usepackage[ruled]{algorithm2e}

\newtheorem{assumption}{Assumption}

\newtheorem{lemma}{Lemma}
\newtheorem{theorem}{Theorem}

\newcommand{\one}{\mathbbm{1}} 
\newcommand{\scallones}{\frac{1}{K}\mathbbm{1}\mathbbm{1}^\mathsf{T}} 
\newcommand{\onekronim}{\frac{1}{K}\left(\mathbbm{1}\transp\otimes I_M\right)} 
\newcommand*\w{\boldsymbol {{\scriptstyle \mathcal{W} }} }
\newcommand*\wb{ \boldsymbol{w} }
\newcommand*\gb{ \boldsymbol{g} }
\newcommand*\sbb{ \boldsymbol{s} }
\newcommand*\hb{ \boldsymbol{h} }
\newcommand*\vb{ \boldsymbol{v} }
\newcommand*\x{ {\scriptstyle \mathcal{X} } }
\newcommand*\g{ \boldsymbol{{\mathscr{G} }} }
\newcommand*\y{ \boldsymbol{{\scriptstyle \mathcal{Y}}} }
\newcommand*\F{ \boldsymbol{ \mathcal{F} } }
\newcommand*\A{\mathcal{A}}
\newcommand*\Ahat{\widehat{\A}}
\newcommand*\transp{^\mathsf{T}}
\newcommand*\grad{\nabla\mathcal{J}}
\newcommand*\gradstoch{\widehat{\grad}}
\newcommand*\xhat{\widehat{\boldsymbol{\x}}}
\newcommand*\what{\widehat{\w}}
\newcommand*\yhat{\widehat{\y}}
\newcommand*\Ihat{\widehat{\mathcal{I}}}
\newcommand*\s{ \boldsymbol{{\scriptstyle \mathcal{S} }} }
\newcommand*\epstau{\epsilon_\tau}
\newcommand*\lbreak{\nonumber\\&\quad}

\newcommand*\centerr{ \widetilde{\wb} }

\newcommand\expect[1]{%
\mathbb{E}\left[#1\right]
}
\newcommand\expectfi[1]{%
\expect{#1 \Bigl|\F_{i-1}}
}
\newcommand\norm[1]{%
\left\lVert#1\right\rVert
}

\begin{document}

\title{Decentralized Learning with Approximate Finite-Time Consensus}

\author{\IEEEauthorblockN{Aaron Fainman and Stefan Vlaski}
\thanks{The authors are with the Department of Electrical and Electronic Engineering, Imperial College London. This work was supported in part by EPSRC Grants EP/X04047X/1 and EP/Y037243/1. Emails: \{aaron.fainman22, s.vlaski\}@imperial.ac.uk.}
}

\maketitle

\begin{abstract}
The performance of algorithms for decentralized optimization is affected by both the optimization error and the consensus error, the latter of which arises from the variation between agents' local models. Classically, algorithms employ averaging and gradient-tracking mechanisms with constant combination matrices to drive the collection of agents to consensus. Recent works have demonstrated that using sequences of combination matrices that achieve finite-time consensus (FTC) can result in improved communication efficiency or iteration complexity for decentralized optimization. Notably, these studies apply to highly structured networks, where exact finite-time consensus sequences are known exactly and in closed form. In this work we investigate the impact of utilizing approximate FTC matrices in decentralized learning algorithms, and quantify the impact of the approximation error on convergence rate and steady-state performance. Approximate FTC matrices can be inferred for general graphs and do not rely on a particular graph structure or prior knowledge, making the proposed scheme applicable to a broad range of decentralized learning settings.
\end{abstract}


\section{Introduction and Related Work}
Consider the problem of decentralized aggregate optimization, where a network of $K$ agents aim to collectively solve:
\begin{align}\label{eq:cons-opt}
   \min_{w\in\mathbb{R}^M}\quad J(w)=\frac{1}{K}\sum_{k=1}^K J_k(w)
\end{align}
over a graph $\mathcal{G}=(\mathcal{V},\mathcal{E})$. Each local objective function \({J_k:\mathbb{R}^M\rightarrow\mathbb{R}}\) is known only to agent $k$ and is defined as the expectation of a local loss \(Q_k(w; \boldsymbol{x}_k) \), i.e., \(J_k(w)=\mathbb{E}Q_k(w; \boldsymbol{x}_k)\) where \( \boldsymbol{x}_{k} \) denotes the random data available to agent \( k \) and the expectation is taken over the distribution of \( \boldsymbol{x}_k \).

Algorithms that solve~\eqref{eq:cons-opt} in a decentralized manner are generally composed of a local optimization step and a social learning step during which agents share their local variables with the agents in their neighbourhood, $\mathcal{N}_{k}$. This mixing step often resembles a consensus iteration of the form:
\begin{align}\label{eq:linear-avg}
    w_{k,i}= \sum_{\ell\in\mathcal{N}_k} a_{\ell k} w_{\ell,i-1}
\end{align}
where \( a_{\ell k} = [A]_{\ell, k} \) are elements of a weighted combination matrix for the graph. \\

Convergence rates and performance bounds of decentralized algorithms typically involve $\lambda_2(A)$, the second largest eigenvalue of $A$~\cite{alghunaim, chen2015, xin21, vlaski21, vlaski_kar_23}. For poorly-connected networks, $\lambda_2(A)$ approaches 1 and the consensus error term will contribute significantly to the learning error. Minimizing \( \lambda_2(A) \) for a given graph topology results in the fastest-mixing combination matrix, which minimizes the upper bound on the consensus error for most decentralized optimization algorithms~\cite{xiao}. Alternative constructions for optimal combination policies that take into account data statistics, such as the Metropolis-Hastings rule~\cite{sayed2014}, have also been considered in the literature.

By utilizing a carefully constructed sequence of time-varying combination weights \( A_{i} \) in~\eqref{eq:linear-avg}, it is possible to achieve exact consensus in a finite number of iterations. These sequences are known as finite-time consensus (FTC) sequences~\cite{Ko_Shi_2009,ko,hendrickx,Shi_Li}. Their defining feature is that the product of matrices over the entire sequence equals the scaled all-ones matrix:
\begin{align}
    A_\tau \cdots A_2 A_1=\scallones
\end{align}
Averaging with FTC sequences takes the form:
\begin{align}
    w_{k,i}=\sum_{\ell\in\mathcal{N}_k} a_{\ell k,i}w_{\ell,i-1}
\end{align}
where \( a_{\ell k, i} \) depends on \( i \) and returns the exact average of agents' initial models \( w_{k, 0} \) in $\tau$ steps. The number of matrices in the sequence, $\tau$, is known as the graph's consensus number and is lower bounded by the graph's diameter and upper bounded by twice the graph's radius~\cite{hendrickx}. An example of an FTC sequence on a hypercube with 4 agents is shown in Fig.~\ref{fig:ftc-ex}.

\renewcommand*{\arraystretch}{1.2}
\begin{center}
\begin{figure}[]
    \centering 
\begin{subfigure}{0.25\linewidth}
  \includegraphics[width=\linewidth]{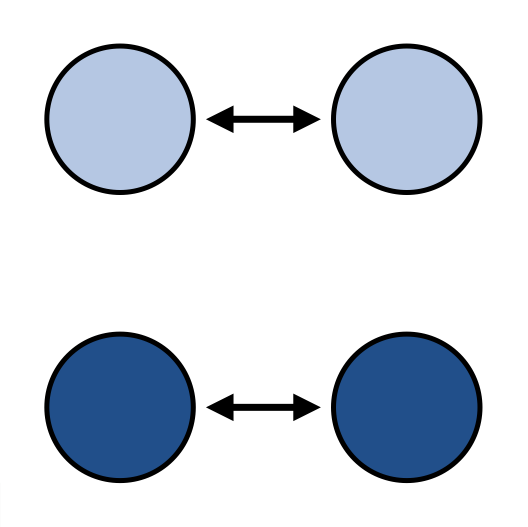}
\end{subfigure}\hspace{38pt}
\begin{subfigure}{0.25\linewidth}
  \includegraphics[width=\linewidth]{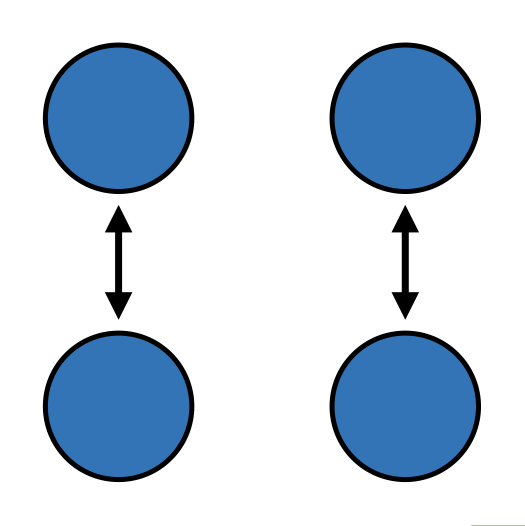}
\end{subfigure}\\
\begin{subfigure}{0.41\linewidth}
  \begin{align*}
      \begin{bmatrix}
          \frac{1}{2} & \frac{1}{2} & 0 & 0 \\
          \frac{1}{2} & \frac{1}{2} & 0 & 0 \\
          0 & 0& \frac{1}{2} & \frac{1}{2}  \\
          0 & 0 & \frac{1}{2} & \frac{1}{2} \\
      \end{bmatrix}
  \end{align*}\normalsize
\end{subfigure}
\begin{subfigure}{0.41\linewidth}
  \begin{align*}
      \begin{bmatrix}
          \frac{1}{2} & 0 & \frac{1}{2} & 0\\
          0 & \frac{1}{2} & 0 & \frac{1}{2} \\
          \frac{1}{2} & 0 & \frac{1}{2} & 0\\
          0 & \frac{1}{2} & 0 & \frac{1}{2} \\
      \end{bmatrix}
  \end{align*}\normalsize
\end{subfigure}
\caption{Finite-time consensus (FTC) sequence for a hypercube with 4 agents. The product of the two matrices equals ${\frac{1}{4}\one\one\transp}$.}
\label{fig:ftc-ex}
\end{figure}\vspace{-25pt}
\end{center}

\vspace{-18pt}FTC sequences have been used as the combination matrices in decentralized algorithms with demonstrable benefits~\cite{Cesar,eusipco,ying-exp}. In decentralized momentum SGD, FTC sequences applied to 1-peer exponential graphs can enable sparser communication without compromising the convergence rate~\cite{ying-exp}. Similar results have been observed in gradient-tracking algorithms~\cite{Cesar}, while faster convergence rates have been demonstrated on other graph families~\cite{eusipco}. Empirically, it has been observed in~\cite{eusipco} that benefits to the performance of gradient-tracking also extend to the case when the sequence of matrices only approximates the scaled all ones matrix. We refer to such sequences of matrices as \emph{approximate} finite-time consensus (FTC) sequences.

Approximate FTC sequences may arise from the numerical methods for calculating the FTC sequences. Closed-form rules for FTC sequences are only known for certain families of graphs~\cite{delvenne, ying-exp, Shi_Li, Cesar}. For arbitrary graphs, provided that the graph topology is known, an eigendecomposition of the adjacency matrix~\cite{kibangou, safavi,sandryhaila} or graph filter design~\cite{coutino, segarra} can be used to find the FTC sequence. Alternatively, the sequences may be learned in a decentralized fashion~\cite{eusipco}. In these and other cases, we may not be provided with an exact FTC sequence. Numerical inaccuracies in the eigendecomposition or an underestimation of $\tau$ may, for example, only yield the approximate sequence. We can quantify the quality of the approximation as:
\begin{align}\label{eq:epsilon}
    \epstau = \norm{A_{\tau}\cdots A_2A_1-\scallones}
\end{align}
with perfect FTC being defined by $\epstau=0$. Current analytical guarantees for the performance of gradient-tracking based algorithms for decentralized optimization apply only to the case of \emph{exact} FTC sequences (i.e., \( \epsilon_{\tau} = 0 \))~\cite{Cesar}. In this work, we develop convergence guarantees allowing for approximate FTC sequences \( \epsilon_{\tau} > 0 \), and clarify its impact on performance.

\section{Analysis}

In this work, we study the performance of the Aug-DGM algorithm~\cite{augdgm} with approximate FTC sequences. The algorithm consists of a coupled recursion between the model variable $\wb_{k,i}$ and an auxiliary variable \( \gb_{k, i} \) that tracks the gradient of the aggregate cost~\eqref{eq:cons-opt} through a dynamic consensus recursion:
\begin{subequations}
\begin{align}
    \wb_{k,i} &= \sum_{\ell\in\mathcal{N}_k} a_{\ell k, i}\left(\wb_{\ell,i-1}-\gb_{\ell,i-1}\right)  \label{eq:aug1_node}\\
    \gb_{k,i} &= \sum_{\ell\in\mathcal{N}_k} a_{\ell k, i}\left(\gb_{\ell,i-1} + \mu\widehat{\nabla J}_{\ell}(\wb_{\ell,i})- \mu\widehat{\nabla J}_{\ell}(\wb_{\ell,i-1}) \right)  \label{eq:aug2_node}
\end{align}
\end{subequations}
where $\mu$ is the step size parameter and $\widehat{\nabla J}_k(\cdot)$ represents a stochastic approximation of the true gradient, $\nabla J_k(\cdot)$. A typical choice of the gradient approximation is \( \widehat{\nabla J}_{k} (\wb_{k, i-1}) \triangleq \nabla Q_k(\wb_{k, i-1}; \boldsymbol{x}_{k, i}) \), but we allow for other constructions such as mini-batches as well. We use bold notation to denote random variables. Motivated by~\cite{Cesar} and deviating from the classical implementation~\cite{augdgm}, we include the step size \( \mu \) in the gradient-tracking recursion~\eqref{eq:aug2_node} rather than the gradient update in~\eqref{eq:aug1_node}. This causes \( \gb_{k, i} \) to estimate \( \mu \nabla J(\cdot) \) rather than \( \nabla J(\cdot) \), and reduces the accumulation of errors in the gradient-tracking recursion. To employ the FTC sequences, the combination matrices are cycled over the sequence, such that, ${A_i=A_{i\%\tau}}$, where \( \% \) denotes the modulo operation.

Recursions~\eqref{eq:aug1_node}-\eqref{eq:aug2_node} can be represented more compactly using network quantities, ${\w_i\triangleq\text{col}\{\wb_{k,i}\}}$, ${\g_i\triangleq\mathrm{col}\{\gb_{k,i}\}}$, and ${\gradstoch(\w_i)\triangleq\text{col}\{\widehat{\nabla J}_k(\wb_{k,i})\}}$:
\begin{subequations}\label{eq:aug}
    \begin{align}
    \w_{i}&=\A_{i}\left(\w_{i-1}-\g_{i-1} \right) \label{eq:aug1}\\
    \g_{i}&=\A_{i}\left(\g_{i-1}+\mu\gradstoch(\w_{i})-\mu\gradstoch(\w_{i-1}) \right)  \label{eq:aug2}
    \end{align}
\end{subequations}
where $\A_i=A_i\otimes I_M$.

Following~\cite{Cesar}, for the convenience of analysis, the coupled recursion~\eqref{eq:aug1}-\eqref{eq:aug2} is transformed using the change of variable ${\y_i\triangleq\g_{i}-\mu\A_i\gradstoch(\w_i)}$ initialized with $\y_0=0$. This removes \( \w_i \) from the update for the tracking variable at time $i$~\cite{Cesar}. The equivalent pair of recursions is:
\begin{subequations}\label{eq:aug-y}
    \begin{align}
        \w_i&=\A_i\w_{i-1}-\A_i\y_{i-1}-\mu\A_i\A_{i-1}\gradstoch(\w_{i-1}) \label{eq:aug-y-w}\\
        \y_i &= \A_i\y_{i-1}-\mu\A_i(I-\A_{i-1})\gradstoch(\w_{i-1}) \label{eq:aug-y-y}
    \end{align}
\end{subequations}

\subsection{Overview}
We begin the analysis by bounding the consensus error in Section~\ref{subsec:cons-err}. This begins by quantifying the disagreement in the agents' local models, ${\what_i\triangleq \Ihat\w_i = \w_i-\one\otimes\wb_{c,i}}$, where ${\Ihat\triangleq(I-\scallones)\otimes I_M}$. The disagreement in the auxiliary variables is measured similarly, ${\yhat_i\triangleq\Ihat\y_i}$. We will consider and bound the joint disagreement ${\xhat_i\triangleq[\what_i\transp,\yhat_i\transp]\transp}$, and term this the consensus error, on the grounds that if $\xhat_i$ is bounded in the mean-square sense, then $\what_i$ must also be bounded.

In Section~\ref{subsec:main-res}, we describe the evolution of the network centroid, defined by ${\wb_{c,i}\triangleq\onekronim \w_{i}=\frac{1}{K}\sum_{k=1}^K \wb_{k,i}}$ and then bound the difference between this and the optimal model, $w^o$. This is combined with the consensus error bound to give the main result in Theorem~\ref{thm:main-bound}.

\subsection{Assumptions}
The analysis is conducted under regularity conditions on the individual and aggregate cost function.
\begin{assumption}[Regularity conditions]\label{assump:regularity}
The aggregate objective function $J(\cdot)$ is strongly-convex:
\begin{align}
    J(y)\geq J(x)+\nabla J(x)\transp(x-y)+\frac{\nu}{2}\lVert y-x\rVert^2 
\end{align}
and the local gradients are Lipschitz smooth:
\begin{align}
    \lVert \nabla J_k(x)-\nabla J_k(y)\rVert \leq \delta\lVert x-y\rVert
\end{align}
\end{assumption}
Additionally, we place a uniform bound on the gradient heterogeneity, which simplifies the analysis by allowing us to bound the consensus error independently of the optimization error in Theorem~\ref{thm:cons-bound}.
\begin{assumption}[Bounded gradient heterogeneity]\label{assump:bounded}
The disagreement in gradients between any two agents, $k$ and $\ell$, is bounded:
\begin{align}
    \left\lVert \nabla J_k(x)-\nabla J_\ell(x) \right\rVert \leq B
\end{align}
\end{assumption}
We impose classical conditions on the gradient noise, which quantifies the quality of the gradient approximation \( \widehat{\nabla J}_k(\wb_{k, i-1}) \).
\begin{assumption}[Gradient noise]\label{assump:gradnoise}
The gradient noise, $\sbb_{k,i}(\wb_{k,i-1})$, defined by:
\begin{align}
    \sbb_{k,i}(\wb_{k,i-1})\triangleq \widehat{\nabla J}_k(\wb_{k,i-1})-\nabla J_k(\wb_{k,i-1})
\end{align}
is unbiased, pairwise-uncorrelated and has a bounded variance conditioned on the filtration $\F_{i-1}$, which contains all randomness up to and including time \( i-1 \):
\begin{subequations}
    \begin{align}
        \expectfi{\sbb_{k,i}(\wb_{k,i-1})} &=0 \\
        \expectfi{\sbb_{k,i}\transp(\wb_{k,i-1}) \sbb_{\ell,i}(\wb_{\ell,i-1})} &=0 \\
        \expectfi{\norm{\sbb_{k,i}(\wb_{k,i-1})}^2} &\leq \sigma_k^2
    \end{align}
The conditionally unbiased and pairwise-uncorrelated assumptions on the gradient noise will extend to the stacked gradient noise vector, ${\s_i(\w_{i-1})\triangleq\text{col}\{\sbb_{k,i}(\wb_{k,i-1})\}}$, while the variance is bounded by:
\begin{align}
    \expectfi{\norm{\s_i(\w_{i-1})}^2} &\leq \sum_{k=1}^K\sigma_k^2 \triangleq \sigma^2
\end{align}
\end{subequations}
\end{assumption}
Finally, we require standard conditions on the combination matrices.
\begin{assumption}[Combination Matrices]\label{assump:comb-mat}
Each combination matrix ${\{A_j\}_{j=1}^{\tau}}$ in the FTC sequence is primitive, doubly-stochastic, and has a spectral radius of 1.
    
\end{assumption}

\subsection{Consensus Error}\label{subsec:cons-err}
The evolution of the consensus error error is found by pre-multiplying~\eqref{eq:aug-y} by ${\Ihat\triangleq (I-\scallones)\otimes I_M}$ to give: 
\begin{subequations}
\begin{align}\label{eq:cons-err}
    \xhat_i=G_i\xhat_{i-1}-\mu \hb_i - \mu \vb_i
\end{align}
where:
\begin{align}
    \Ahat_i &\triangleq \Ihat\A_i \\
    G_i &\triangleq \begin{bmatrix}
        \Ahat_i & -\Ahat_i \\
        0 & \Ahat_i 
    \end{bmatrix} \\
    \hb_{i} &\triangleq \begin{bmatrix}
\A_{i}\A_{i-1}\Ihat\grad(\w_{i-1}) \\
\A_{i} (I-A_{i-1})\Ihat\grad(\w_{i-1})
\end{bmatrix} \\
\vb_{i} &\triangleq \begin{bmatrix}
    \A_{i}\A_{i-1}\Ihat\s_{i}(\w_{i-1}) \\
    \A_{i}(I-\A_{i-1})\Ihat\s_{i}(\w_{i-1})
\end{bmatrix}
    \end{align}
\end{subequations}\normalsize 

In~\eqref{eq:cons-err} the evolution of the consensus error is defined in relation to the previous iterate, $\xhat_{i-1}$. In order to make use of the approximate FTC property we are required to consider the consensus error over at least $\tau$ iterations. We do this by repeatedly substituting~\eqref{eq:cons-err} for $\xhat_{i-1}$ and then $\xhat_{i-2}$ and so forth, so that we obtain:
\begin{align}
    \xhat_{i} &= G_{i:i-1}\xhat_{i-2}-\mu G_i\hb_{i-1} - \mu \hb_i -\mu G_i\vb_{i-1} - \mu \vb_i \nonumber \\
     &= G_{i:i-2}\xhat_{i-3}-\mu G_{i:i-1}\hb_{i-2}-\mu G_i\hb_{i-1}- \mu \hb_i \lbreak-\mu G_{i:i-1}\vb_{i-2}-\mu G_i\vb_{i-1}- \mu \vb_i \nonumber\\
     &= \cdots \nonumber\\
    \xhat_i &= G_{i:m\tau+1}\xhat_{m\tau}-\mu\sum_{\mathclap{j=m\tau+1}}^{i-1}G_{i:j+1}\hb_{j}-\mu\hb_i\lbreak-\mu\sum_{\mathclap{j=m\tau+1}}^{i-1}G_{i:j+1}\vb_{j} -\mu\vb_i \label{eq:cons-err-mtau}
    \end{align}
where $m\triangleq\left\lfloor\frac{i}{\tau}\right\rfloor-1$ and we introduced the short-hand notation \( G_{i:j} \triangleq G_i G_{i-1} \cdots G_{j} \). Iterating over the recursions until $\xhat_{m\tau}$ ensures that there are between $\tau$ and $2\tau-1$ iterations that have been considered in the recursion.

Prior to bounding the consensus error in~\eqref{eq:cons-err-mtau} we state three lemmas necessary to bound the expression, the proofs of which follow from Assumptions 1 through 4 along with Jensen's inequality.

\begin{lemma}[Contraction Rate]
   For $m\triangleq\left\lfloor\frac{i}{\tau}\right\rfloor-1$:
   \begin{align}
       \norm{G_{i:m\tau+1}} \leq \epstau
   \end{align}
\end{lemma}
with $0\leq\epstau<1$ defined in~\eqref{eq:epsilon} which follows since the spectral norm of a block triangular matrix equals the maximum spectral norm of its diagonal blocks~\cite{simovici}, i.e. $\lVert\Ahat_{i:m\tau+1}\rVert=\lVert\Ahat_{i\%\tau:1}\rVert\leq\epstau$.\\

\begin{lemma}\label{lemma:h_i}
    Under Assumptions~\eqref{assump:regularity},~\eqref{assump:bounded} and~\eqref{assump:comb-mat}, $\hb_i$ is bounded by:
    \begin{align}
        \expectfi{\norm{\hb_i}^2 } &\leq 18\delta^2\norm{\xhat_{i-1}}^2+9KB^2
    \end{align}
\end{lemma}

\begin{lemma}\label{lemma:v_i}
    Under Assumptions~\eqref{assump:regularity},~\eqref{assump:gradnoise} and~\eqref{assump:comb-mat}, $\vb_i$ is bounded by:
    \begin{align}
        \expectfi{\norm{\vb_i}^2 } &\leq 3\sigma^2
    \end{align}
\end{lemma}

These intermediate lemmas allow us to establish that the consensus error decays every \( \tau \) iterations up to constant terms which are proportional to \( \mu^2 \).
\begin{theorem}\label{thm:cons-bound}
    Under Assumptions~\eqref{assump:regularity}-\eqref{assump:comb-mat}, and for ${0<\epstau<\frac{2}{3}}$ and $\mu\leq\frac{1}{12\delta^2}\sqrt{\frac{1-\epstau}{\tau(2\tau-1)(1+\epstau)}}$, the consensus error in~\eqref{eq:cons-err} is bounded by:
    \begin{subequations}\label{eq:cons-bound-final}\begin{align}
        \mathbb{E}\norm{\xhat_i}^2 &\leq \left(\frac{3}{8}\epstau(2+3\epstau)\right)^{\left\lfloor\frac{i}{\tau}\right\rfloor}\mathbb{E}\norm{\widehat{x}_0}^2 +\frac{432\mu^2 K\tau^2}{(1-\epstau)^2}B^2\nonumber \\&\qquad +\frac{144\mu^2\tau^2}{(1-\epstau)^2}\sigma^2
    \end{align}
    \normalsize and for $\epstau=0$:
    \begin{align}
    \mathbb{E}\norm{\xhat_i}^2 &\leq 27\mu^2\tau(2\tau-1)KB^2+3\mu^2\tau(2\tau-1)\sigma^2
    \end{align}\end{subequations}\normalsize
    \textit{Proof:} Omitted due to space limitations. 
\end{theorem}

\vspace{5pt}
\subsection{Main Result}\label{subsec:main-res}
We begin with the behaviour of the centroid error, which is found by premultiplying~\eqref{eq:aug-y-w} by $\onekronim$. The centroid for the gradient-tracking variable $\y_i$ is 0, which follows from~\eqref{eq:aug-y-y} since \(\onekronim(I-\A_{i-1})=0 \) and thus:
\begin{align}
        \onekronim\y_{i} &= \onekronim\y_{i-1} \nonumber \\
        &= \onekronim\y_{0} = 0
\end{align}
Letting \(\widetilde{\wb}_i \triangleq w^o - \wb_i\), the dynamics of the centroid error are then described by:
\begin{align}\label{eq:centerr}
    \centerr_{c,i} &= \centerr_{c,i-1}-\frac{\mu}{K}(\one\transp\otimes I_M)(\grad(\w_{i-1})+\s_i(\w_{i-1}))
\end{align}\normalsize
    For $\mu\leq\frac{\nu}{\delta^2}$, the centroid error in~\eqref{eq:centerr} is bounded by:
    \begin{align}\label{eq:cent-bound-fin}
        \mathbb{E}\norm{\centerr_{c,i}}^2 &\leq \sqrt{1-2\mu\nu+\mu^2\delta^2}\mathbb{E}\norm{\centerr_{c,i-1}}^2+\frac{2\mu\delta^2}{\nu K}\mathbb{E}\norm{\xhat_{i-1}}^2\lbreak+\frac{\mu^2}{K}\sigma^2
    \end{align}
Iterating the result this recursion and applying Theorem 1, we find the following bound on the mean-squared deviation.

\begin{theorem}\label{thm:main-bound}
    Under Assumptions~\eqref{assump:regularity}-\eqref{assump:comb-mat}, with  ${0<\epstau<\frac{3}{5}}$ and ${\mu\leq\min\left(\frac{\nu}{\delta^2},\frac{1}{12\delta^2}\sqrt{\frac{1-\epstau}{\tau(2\tau-1)(1+\epstau)}}\right)}$, the error at every $\tau^\mathrm{th}$ iteration is bounded by:
    \begin{subequations}
        \begin{align}
            \mathbb{E}\norm{\centerr_{c,\ell\tau}}^2 &\leq \gamma_1^i\norm{\centerr_{c,0}}^2+\frac{\beta_1\mu\delta^2}{\nu K}\gamma_3^i\norm{\xhat_0}^2 \nonumber \lbreak+ \frac{\beta_2\mu^2\delta^2\tau}{\nu^2(1-\epstau)^2}B^2 + \frac{\beta_3\mu^2\delta^2\tau^2}{\nu^2 K(1-\epstau)^2}\sigma^2 \lbreak +\frac{2\mu}{\nu K}\sigma^2
        \end{align}\normalsize
        and for $\epstau=0$:
        \begin{align}
            \mathbb{E}\norm{\centerr_{c,\ell\tau}}^2 &\leq \gamma_1^i\norm{\centerr_{c,0}}^2+ \frac{\beta_4\mu^2\delta^2\tau^2}{\nu^2}B^2 + \frac{\beta_5\mu^2\delta^2\tau^2}{\nu^2 K}\sigma^2 \lbreak +\frac{2\mu}{\nu K}\sigma^2
        \end{align}\normalsize
        where: 
        \begin{align*}
        &\gamma_1=\sqrt{1-2\mu\nu+\mu^2\delta^2},\quad \gamma_2 = \sqrt[\leftroot{-3}\uproot{3}\tau]{\frac{3}{8}\epstau(2+3\epstau)}\\
        &\gamma_3=\max\left(\gamma_1, \gamma_2\right),
        ~\beta_1 = \frac{2}{\lvert \gamma_1-\gamma_2 \rvert}, ~\beta_2=1728,~ \beta_3=576\\
        &\beta_4=108, \quad \beta_5=24
        \end{align*}
    \end{subequations}
\end{theorem}

\section{Simulations and Discussion}
\begin{center}
\begin{figure}[]
    \centering 
\begin{subfigure}{0.85\linewidth}
  \includegraphics[width=\linewidth]{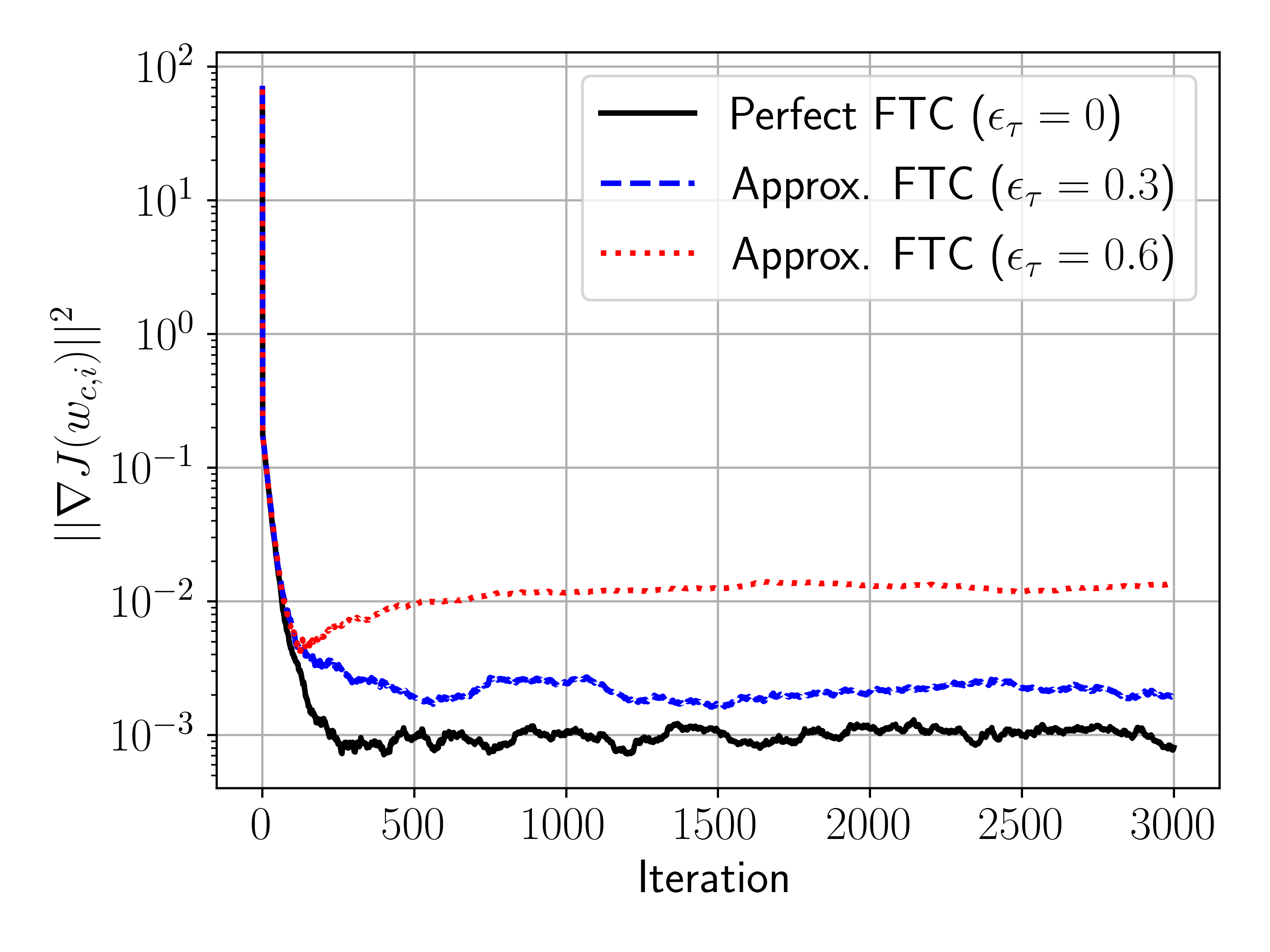}
\end{subfigure}
\caption{Approximate FTC for a logistic regression problem on a path graph.}
\label{fig:log-reg}
\end{figure}
\end{center}

We demonstrate the numerical results on a binary classification problem, with labels $\gamma_{k,n}\in\{-1,1\}$ and features, $h_{k,n}\in\mathbb{R}^M$. Each agent $k$ employs the logistic cost function: 
\begin{align}\label{eq:log-cost}
    J_k(w)=\frac{\rho}{2}\norm{w}^2+\frac{1}{N}\sum_{n=1}^N\ln\left(1+e^{-\gamma_{k,n}h_{k,n}\transp w}\right)
\end{align}
with the number of samples $N=15$, the number of features $M=10$, $\rho=0.01$, and $\mu=0.1$. The stochastic gradient of the empirical logistic cost in~\eqref{eq:log-cost} is computed by selecting a random sample ${1\leq\boldsymbol{n}_i\leq N}$ at each iteration $i$ and evaluating its gradient. The graph used is a path graph with 16 agents and $\tau=15$. Results are shown in Fig.~\ref{fig:log-reg}.

\vspace{-25pt}The figure indicates that increasing $\epstau$ is detrimental to performance. The steady state error increases with larger values of $\epstau$, which matches the prediction in Theorem~\eqref{thm:main-bound}. Increasing $\epstau$ causes the $B^2$ and $\sigma^2$ terms to grow from the $1-\epstau$ factor in the denominator. This is analogous to having a mixing rate close to 0 in the standard Aug-DGM bound, which depends on \( \mathcal{O}\left(\frac{\lambda_2(A)^2}{1-\lambda_2(A)}\right) \). 

\begin{center}
\begin{figure}[]
    \centering 
\begin{subfigure}{0.9\linewidth}
  \includegraphics[width=0.85\linewidth]{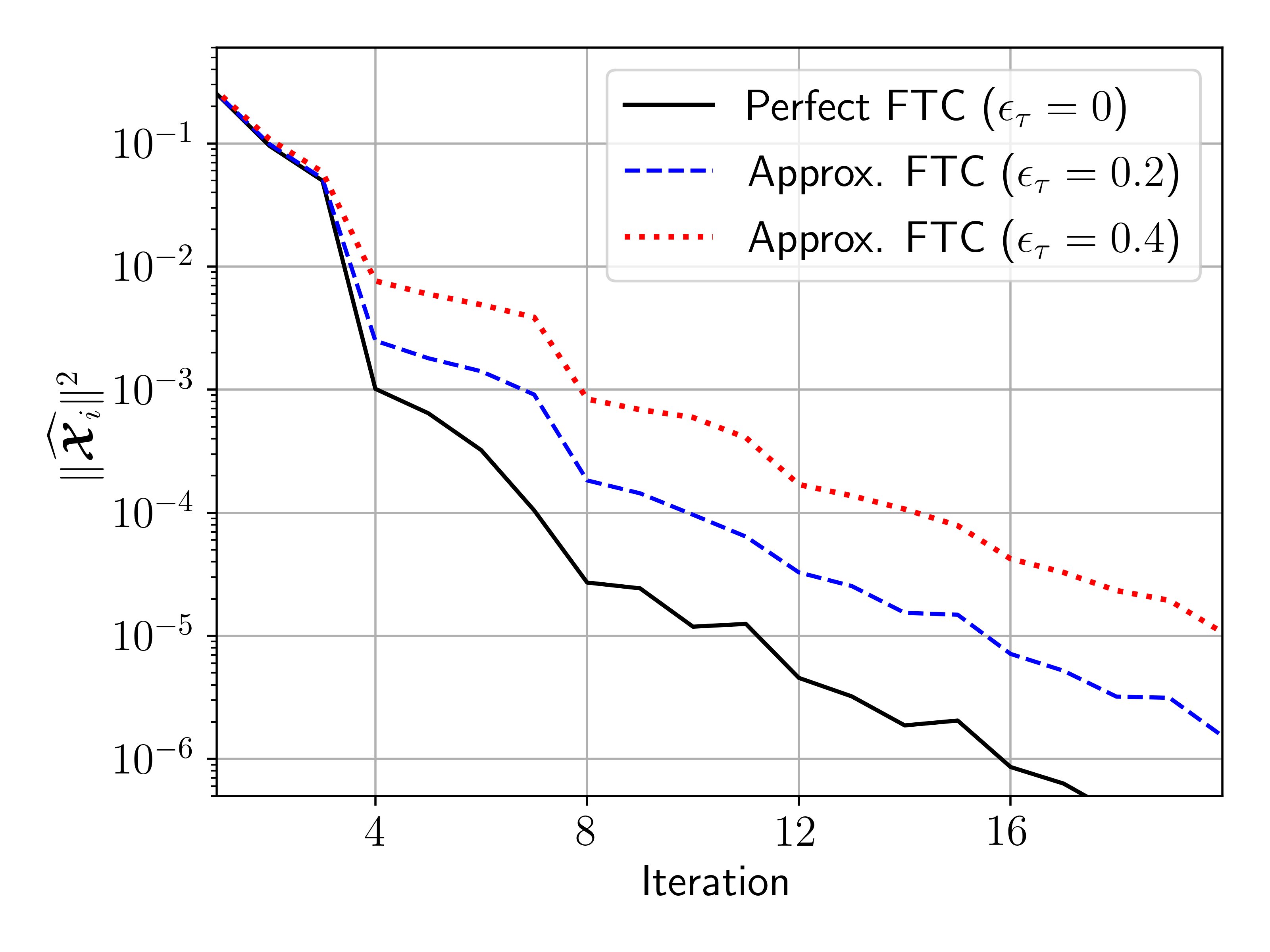}
\end{subfigure}
\caption{Consensus error for a graph with $\tau=4$. The error shows a periodic decrease every 4 iterations. Smaller values of $\epstau$, corresponding to better approximations of the FTC, lead to larger decreases in the error.}
\label{fig:cons-err}
\end{figure}
\end{center}

A larger $\epstau$ also slows down the reduction in the initial consensus error, $\norm{\xhat_0}^2$. We demonstrate this in Fig.~\ref{fig:cons-err} for a hypercube with $\tau=4$ and with the same problem parameters used previously. The consensus bound in Theorem~\ref{thm:cons-bound} depends on $\mathcal{O}(\epstau)^{\left\lfloor\frac{i}{\tau}\right\rfloor}\norm{\xhat_0}^2$, indicating a decrease in the consensus error every $\tau$ iterations. The magnitude of this decrease diminishes with increasing $\epstau$, as demonstrated in Fig.~\ref{fig:cons-err}.

\vspace{-25pt}Increasing $\tau$ also worsens the performance. Both the consensus and centroid error bounds include an $\mathcal{O}(\tau^2)$ term, consistent with the bound in~\cite{Cesar}. Higher values of $\tau$ cause agents' local models to drift from one another because individual combination matrices in the FTC sequence may lack strong connectivity, resulting in a mixing rate of one (see, for example, Fig.~\ref{fig:ftc-ex}). Effective averaging is achieved only over the entire FTC sequence. For larger $\tau$, this leads to greater model drift among agents, which necessitates a smaller step size to compensate, thereby slowing convergence. 

This effect is demonstrated in Fig.~\ref{fig:diff-tau} for the same logistic regression problem but under the deterministic, perfect FTC setting (${\epstau=0,~\sigma^2=0}$). Graphs with $K=16$ agents and different values of $\tau$ have been used, with the step size tuned in each case to give the highest rate of convergence. The results illustrate the performance drawbacks from higher values of $\tau$. The results also demonstrate that that for certain graphs it may be desirable to underestimate $\tau$, trading off the exact consensus sequence for improved performance.

\begin{center}
\begin{figure}[]
    \centering 
\begin{subfigure}{0.9\linewidth}
  \includegraphics[width=0.85\linewidth]{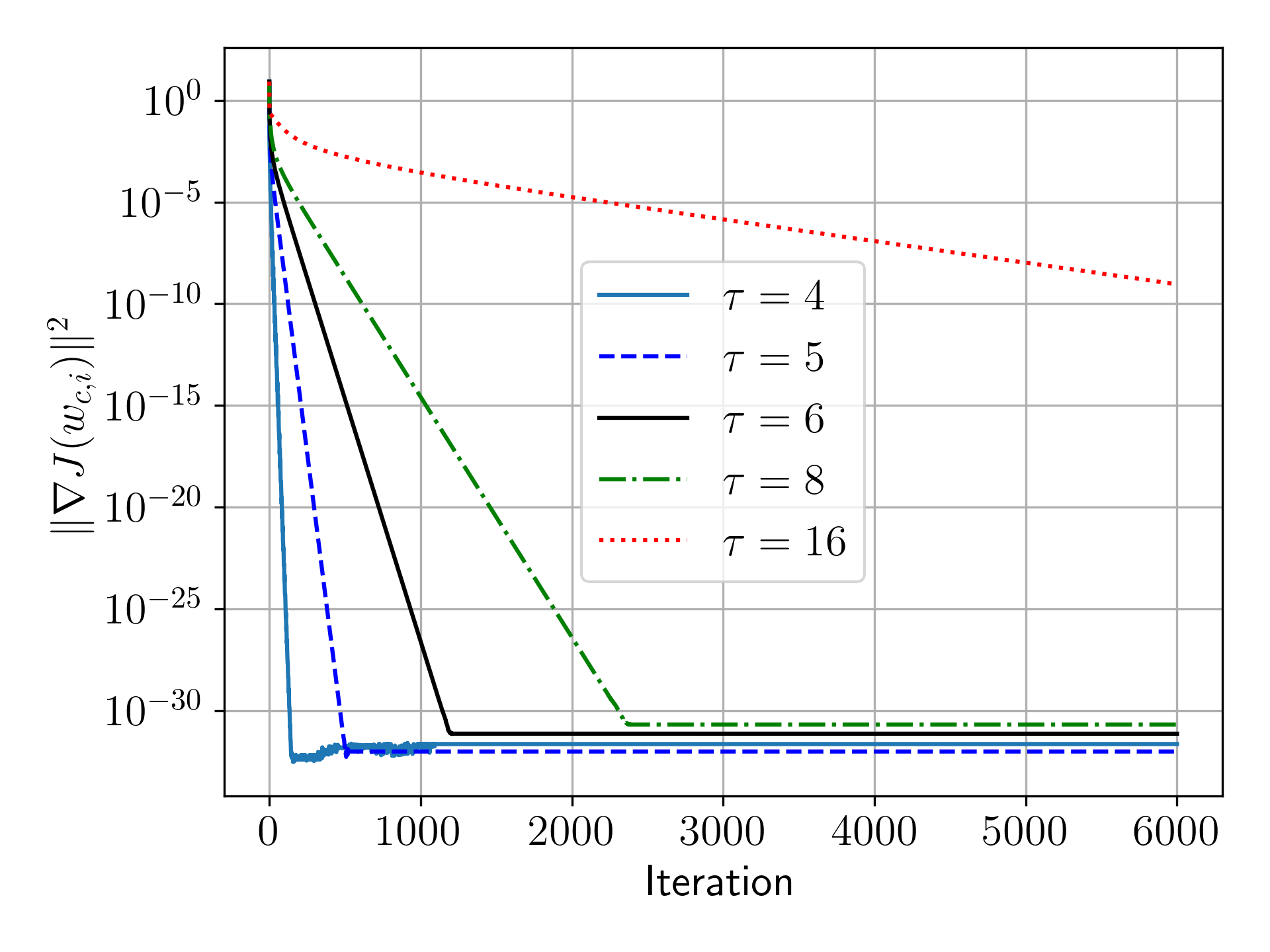}
\end{subfigure}
\caption{Performance comparison on graphs where $K=16$ but $\tau$ ranges in value.}
\label{fig:diff-tau}
\end{figure}\vspace{-20pt}
\end{center}

\section{Conclusion}
We have considered the effect of approximate FTC sequences, which arise from numerical methods for finding FTC sequences. Despite not fully satisfying the exact FTC property, approximate FTC sequences can still provide performance benefits to gradient-tracking algorithms. The bound we have derived predicts better performance the more closely the sequence approximates the scaled all-ones matrix, which matches the simulation results. Both theoretical and numerical results also demonstrate worse performance when increasing the consensus number, $\tau$, demonstrating the utility of FTC sequences for certain families of graphs.

\bibliographystyle{IEEEtran}
\bibliography{References}

\end{document}